\documentclass[]{article}
\usepackage{amsmath}
\usepackage{amssymb}
\usepackage{graphics}
\usepackage{graphicx}
\usepackage{epsfig}
\usepackage{dcolumn}
\usepackage{bm}
\usepackage{lineno}
\usepackage{multirow}
\usepackage{ragged2e}

\title{\bf Effect of charging-up on the uniformity of a single mask triple GEM detector}
\date{}
\begin{document}
	\maketitle
	\vspace*{-1cm}
	\centering
	{
		\author{S.~Chatterjee\footnote[1]{Corresponding Author:
				sayakchatterjee896@gmail.com, sayakchatterjee@jcbose.ac.in},}
		\author{A.~Sen,}
		\author{S.~Das,}
		\author{S.~Biswas}
	}
	\vspace*{0.5cm}
	
	{Department of Physics and Centre for Astroparticle Physics and Space Science (CAPSS), Bose Institute, EN-80, Sector V, Kolkata-700091, India}
	
	\vspace*{0.5cm}
	\centering{\bf Abstract}
	\justify
	The Gas Electron Multiplier~(GEM) detector is one of the advanced members of the Micro Pattern Gas Detector~(MPGD) family, used in High Energy Physics~(HEP) experiments as a tracking device due to its high rate handling capability and good spatial resolution. The uniformity in the performance of the detector is an essential criterion for any tracking device. 
	
	The presence of the dielectric medium~(Kapton) inside the active volume of the GEM chamber changes its behaviour when exposed to external irradiation. This phenomenon is known as the charging-up effect. In this article, the uniformity in terms of gain, energy resolution and count rate of a Single Mask (SM) triple GEM chamber of dimension 10~cm~$\times$~10~cm are reported for both the charged-up and uncharged GEM foils. 
	
	\vspace*{0.25cm}

\section{Introduction}\label{intro}

The Gas Electron Multiplier (GEM) detector is one of the most advanced members of the Micro Pattern Gaseous Detector (MPGD) family~\cite{sauli_GEM}. Due to its high rate handling~($\sim$~1~MHz/mm$^2$) capability and good spatial resolution~($\sim$~70~$\mu$m), GEM is being used in many High Energy Physics~(HEP) experiments as a tracking device where high particle rates are foreseen~\cite{gem_review,ketzer,CMS_upgrade,alice_upgrade,cbm_detector_system}. GEM foil, the heart of the detector, consists of 50~$\mu$m thick Kapton film with 5~$\mu$m Copper cladding on both sides of the Kapton. A large number of holes are pierced in the 60~$\mu$m thin Copper cladded Kapton foil using the photo-lithographic technique. Depending on the technique used, the GEM foils are classified as either Double Mask (DM) or Single Mask (SM) GEM~\cite{GEM_foil}. The GEM foil gets charged up due to the accumulation of the charges on the surface of the foil after amplification inside the GEM holes because of the very high electric field strengths~($\mathcal{O}~({80~kV/cm})$). The behaviour of the detector changes with time when exposed to external irradiation due to the accumulation of the charges on the Kapton surface inside the active volume of the chamber. This phenomenon is known as the charging-up effect~\cite{charging_up_philip,charging_up_azmoun,charging_up_alfonsi}. The charging-up phenomena have been investigated for both the DM and SM triple GEM chambers and reported earlier~\cite{s_chatterjee_charging_up_1,s_chatterjee_charging_up_2}. The uniformity in performance over the entire detector active area is one of the most important criteria for any tracking device. 
Uniformity studies of the performance of GEM chamber prototypes have been performed and reported earlier by various groups~\cite{uniformity_1,uniformity_2,uniformity_3}. 

In this article, the study of the uniformity in the performance of a SM triple GEM chamber of dimension 10~cm~$\times$~10~cm in terms of the gain, energy resolution and count rate with Ar/CO$_{2}$ gas mixture in 70/30 volume ratio is reported considering both charging-up phenomena of GEM foils and also without considering it. The details of the experimental setup are discussed in section~\ref{set_up}. The results are presented in section~\ref{res} and the findings are summarized in section~\ref{summary}.

\section{Detector description and experimental setup}\label{set_up}

\begin{figure}[htb!]
\begin{center}
	\includegraphics[scale=0.40]{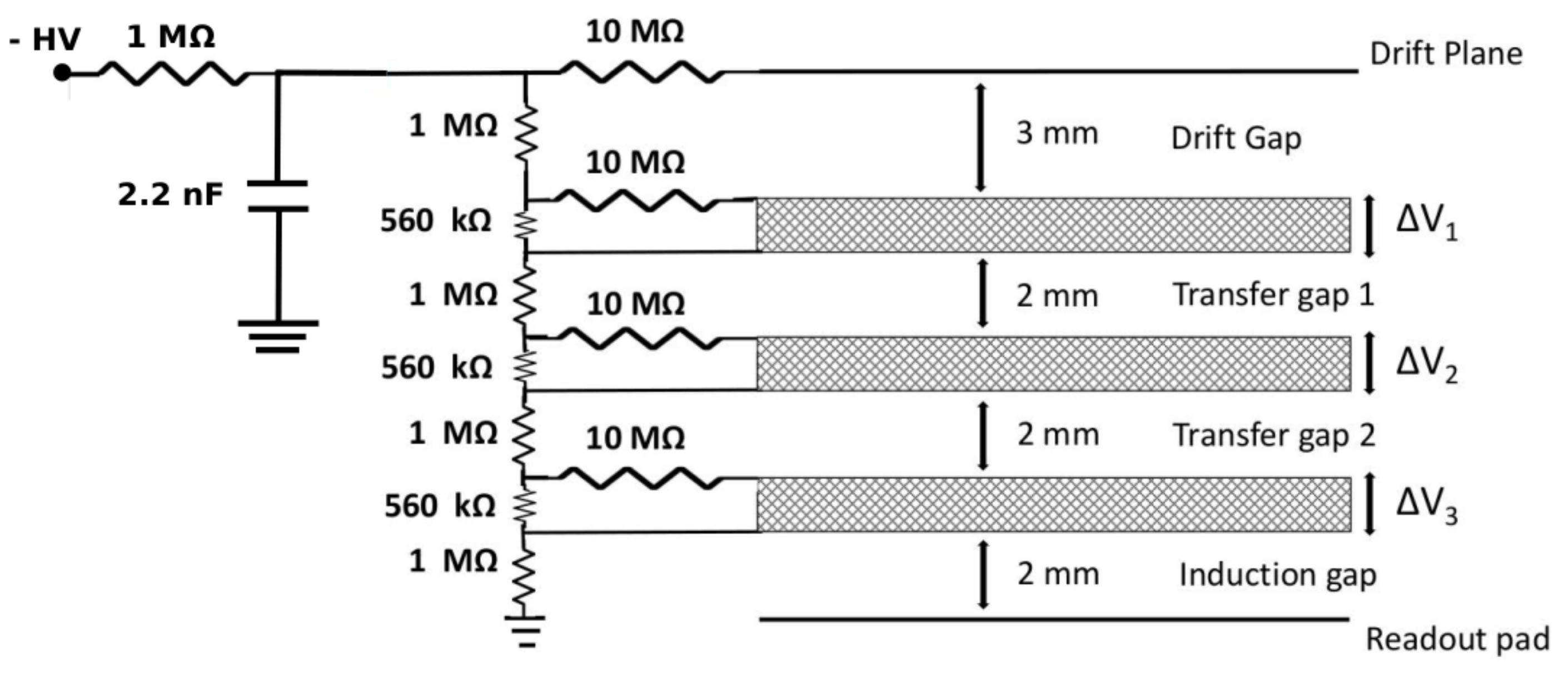}
	\caption{\label{setup} Schematic of the HV distribution through the resistive chain to different planes of the SM triple GEM detector. A HV filter is used between the HV line and the resistive chain~\cite{s_chatterjee_charging_up_2}. }\label{fig1}
\end{center}
\end{figure}

\begin{figure}[htb!]
\begin{center}
	\includegraphics[scale=0.45]{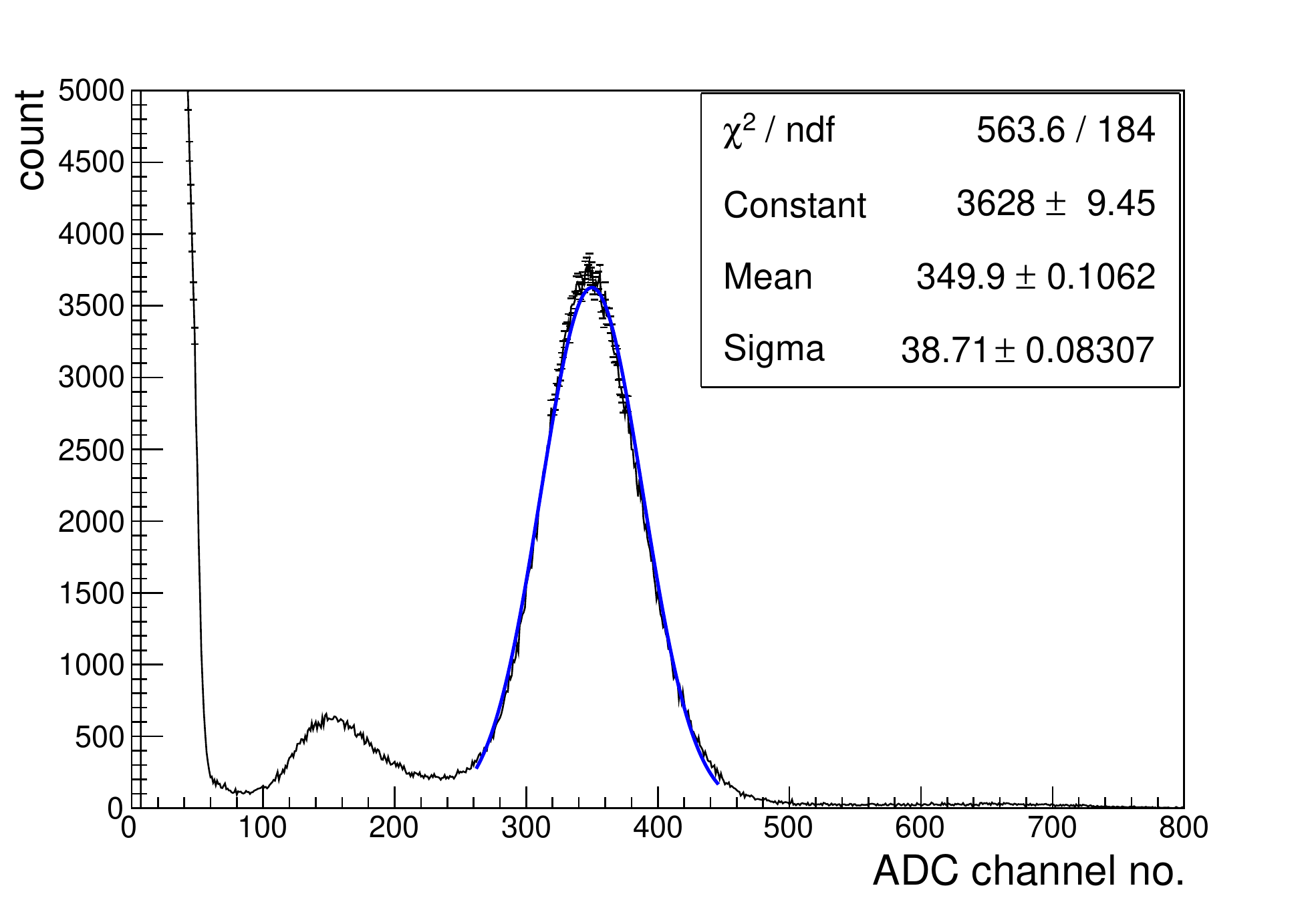}
	\caption{Typical Fe$^{55}$ spectra at - 5075V. The $\Delta V$ across each of the GEM foil is $\sim$~402.7 V (colour online).
	}\label{fig3}
\end{center}
\end{figure}

A SM triple GEM chamber of dimension 10~cm~$\times$~10~cm is used with Ar/CO$_2$ gas mixture in 70/30 volume ratio and in continuous flow mode. The drift gap, transfer gaps and induction gap of the chamber are kept at 3~mm, 2~mm and 2~mm respectively. The GEM foils are biased using a resistive chain network as shown in Fig.~\ref{fig1}. 
A filter circuit is used in between the resistor chain and the input high voltage (HV) to remove the ac components present in the HV line. The readout pad is consist of 256 X and 256 Y tracks. For this study, the signal is taken using sum-up boards instead of the individual readout from the XY tracks. Four sum-up boards are used for this detector. The signal from the sum-up board is fed to a charge sensitive preamplifier having a gain of 2 mV/fC and shaping time of 300~ns~\cite{preamplifier}. The output of the preamplifier is put in a linear Fan-In Fan-Out (FIFO) module. One output from the FIFO is fed to a Multi Channel Analyzer (MCA) to store the X-ray spectra on the desktop. Another output from the FIFO is fed to a Single Channel Analyzer (SCA), the output of which above the
noise threshold is counted using the NIM scaler. A Fe$^{55}$ X-ray source having characteristic energy of 5.9 keV is used to irradiate the chamber.
The typical Fe$^{55}$ X-ray spectrum is shown in Fig.~\ref{fig3} for HV of -5075~V, corresponding to $\Delta$V of $\sim$~402.7~V across each of the GEM foils.
The gain and energy resolution of the chamber is calculated by fitting the 5.9~keV X-ray peak with a Gaussian distribution. The detailed method of gain and energy resolution calculation is discussed in Ref.~\cite{s_chatterjee_charging_up_1, s_roy_gain_calculation}. 

\section{Results}\label{res}
The 10~cm~$\times$~10~cm active area of the SM triple GEM chamber is divided into 4~$\times$~4 regions. A collimator, made of G-10 material, having a diameter of 8~mm and a height of 32~mm is used to irradiate the chamber using the Fe$^{55}$ X-ray source. To measure the uniformity in the characteristics of the chamber, the source is placed on the top of the G-10 collimator and the collimator is placed at the respective positions on the active area of the chamber. 

The data taking is done using two different methodologies. In the first case, the HV is kept ON for $\sim$~60 minutes before starting the measurement and the data taking is started as soon as the source is placed on the chamber. 
\begin{figure}[htb!]
\begin{center}
	\includegraphics[height=5.50cm, width=6.0cm]{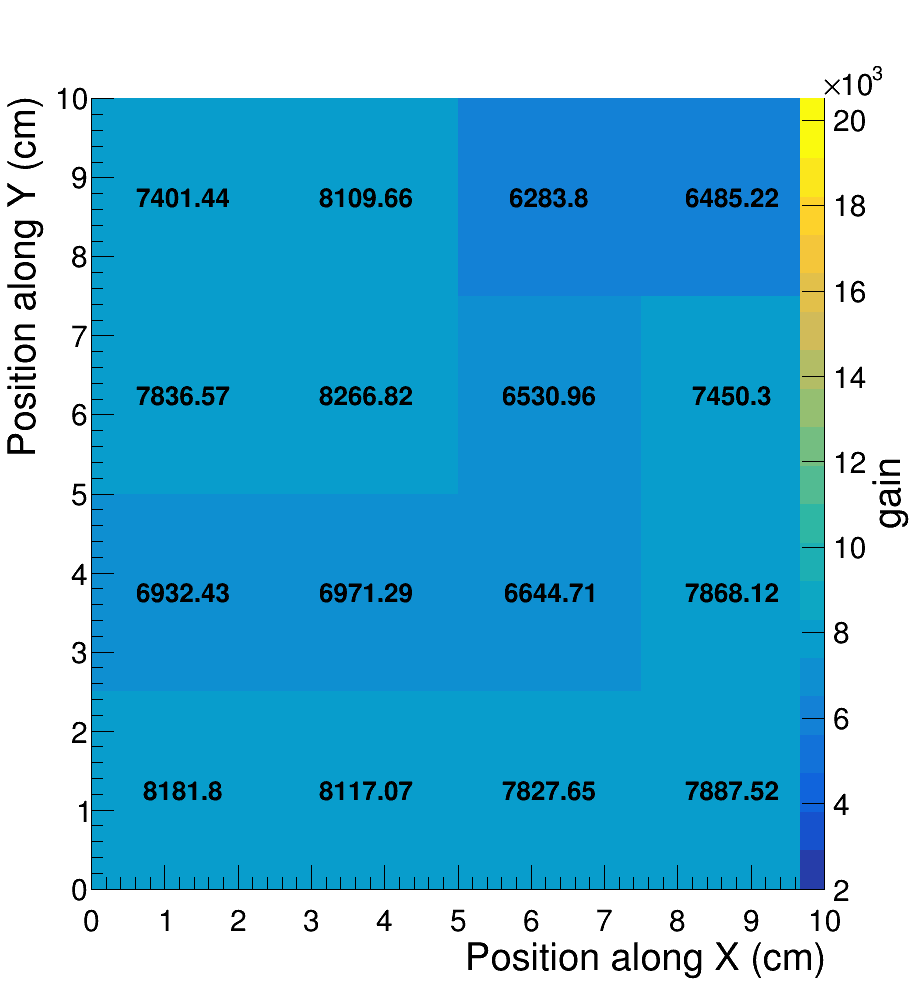}
	\includegraphics[height=5.50cm, width=6.0cm]{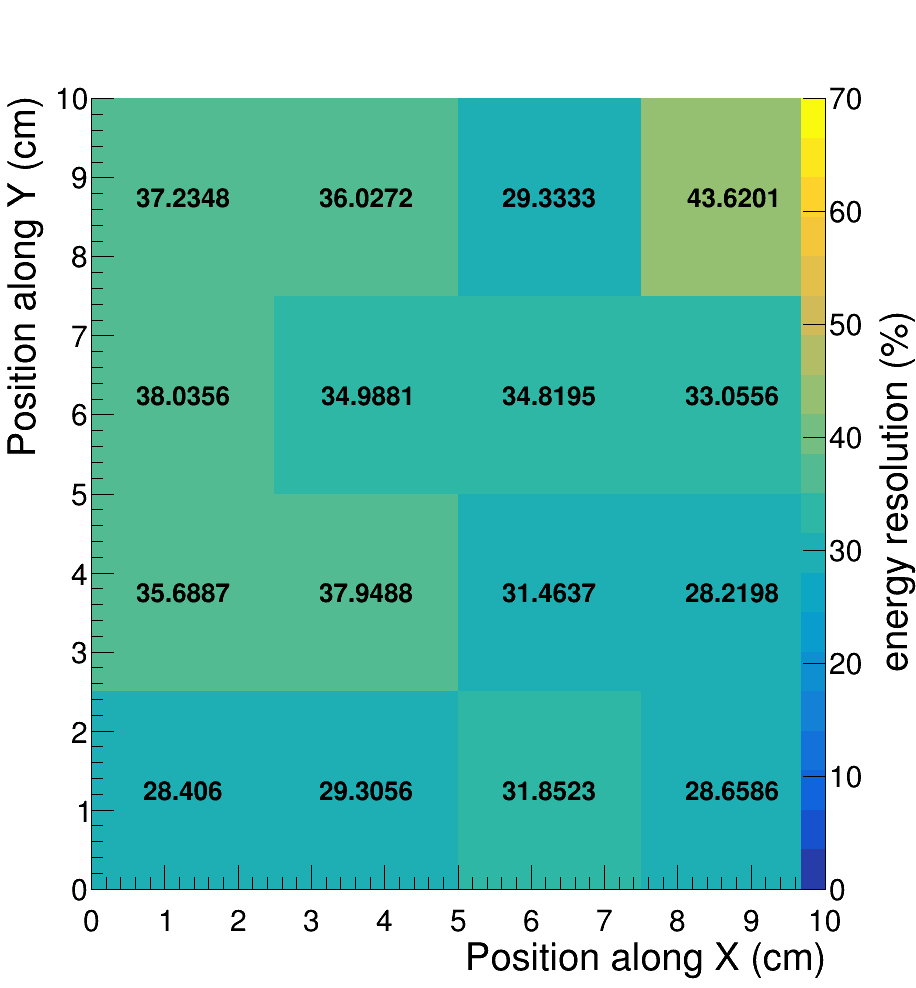}
	\includegraphics[height=5.50cm, width=6.0cm]{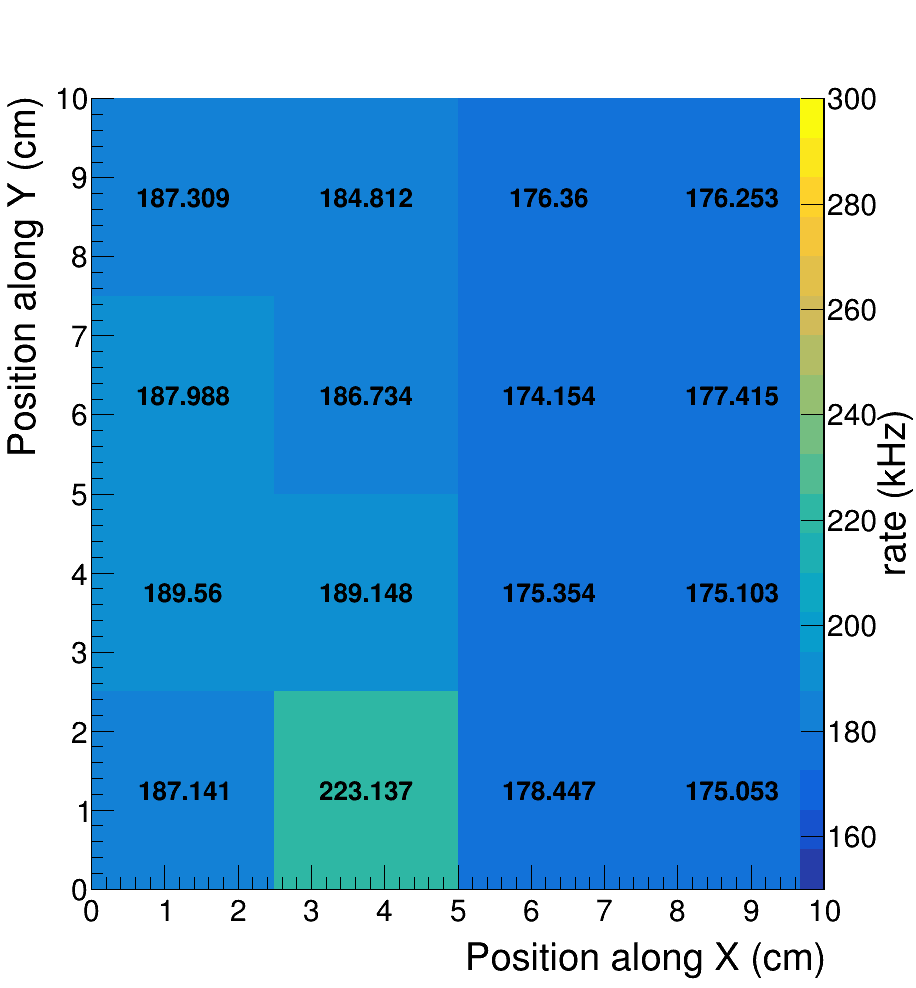}		
	\caption{Variation of gain (top), energy resolution (middle) and count rate (bottom) over the scanned 10~cm~$\times$~10~cm area of the SM triple GEM chamber at a HV of -5075~V. The $\Delta V$ across each of the GEM foil is $\sim$~402.7 V (colour online). }\label{fig4}
\end{center}
\end{figure}
The X-ray energy spectra are recorded for 1 minute and then the source along with the collimator is moved to the next position manually. As a result, the foil does not get sufficient time to get charged up. Therefore, the results we get are essentially with the uncharged GEM foils. In Fig.~\ref{fig4}, the variation in gain, energy resolution and count rate over the scanned area is shown at a $\Delta$V of $\sim$~402.7~V across each of the GEM foils. In Fig.~\ref{fig5}, the distribution of gain, energy resolution and count rate is shown.
\begin{figure}[htb!]
\begin{center}
	\includegraphics[height=5.10cm, width=6.0cm]{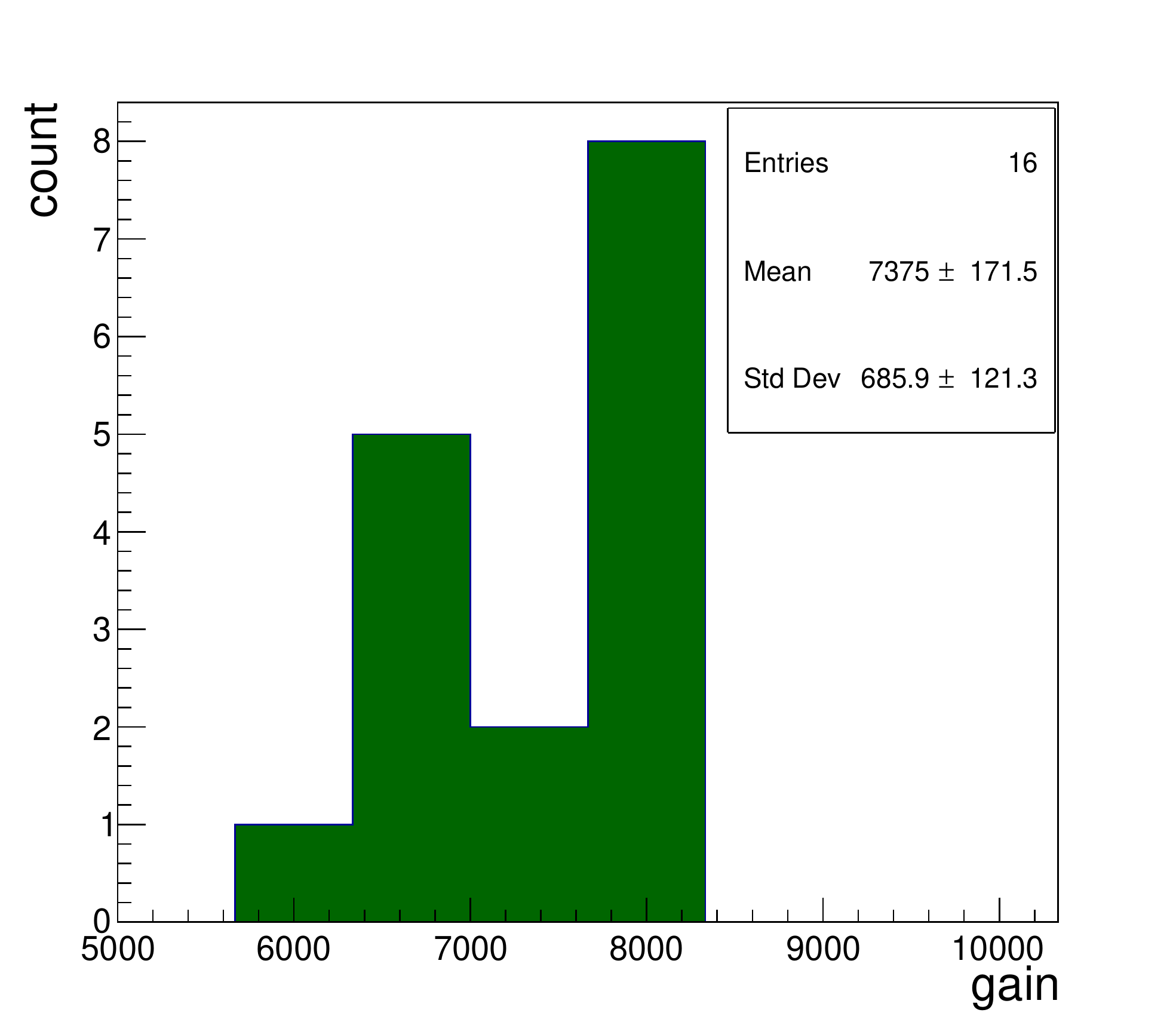}
	\includegraphics[height=5.10cm, width=6.0cm]{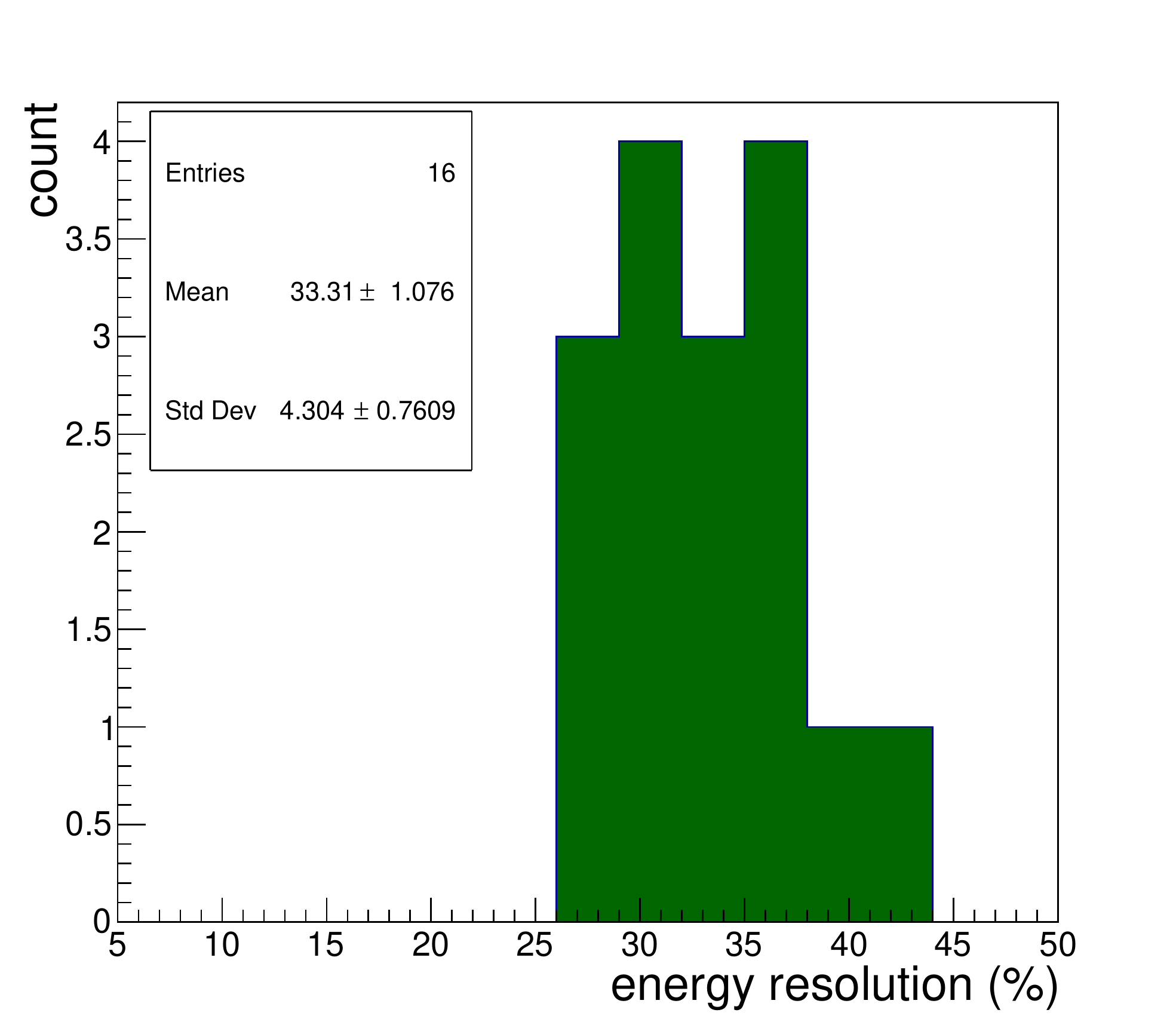}
	\includegraphics[height=5.10cm, width=6.0cm]{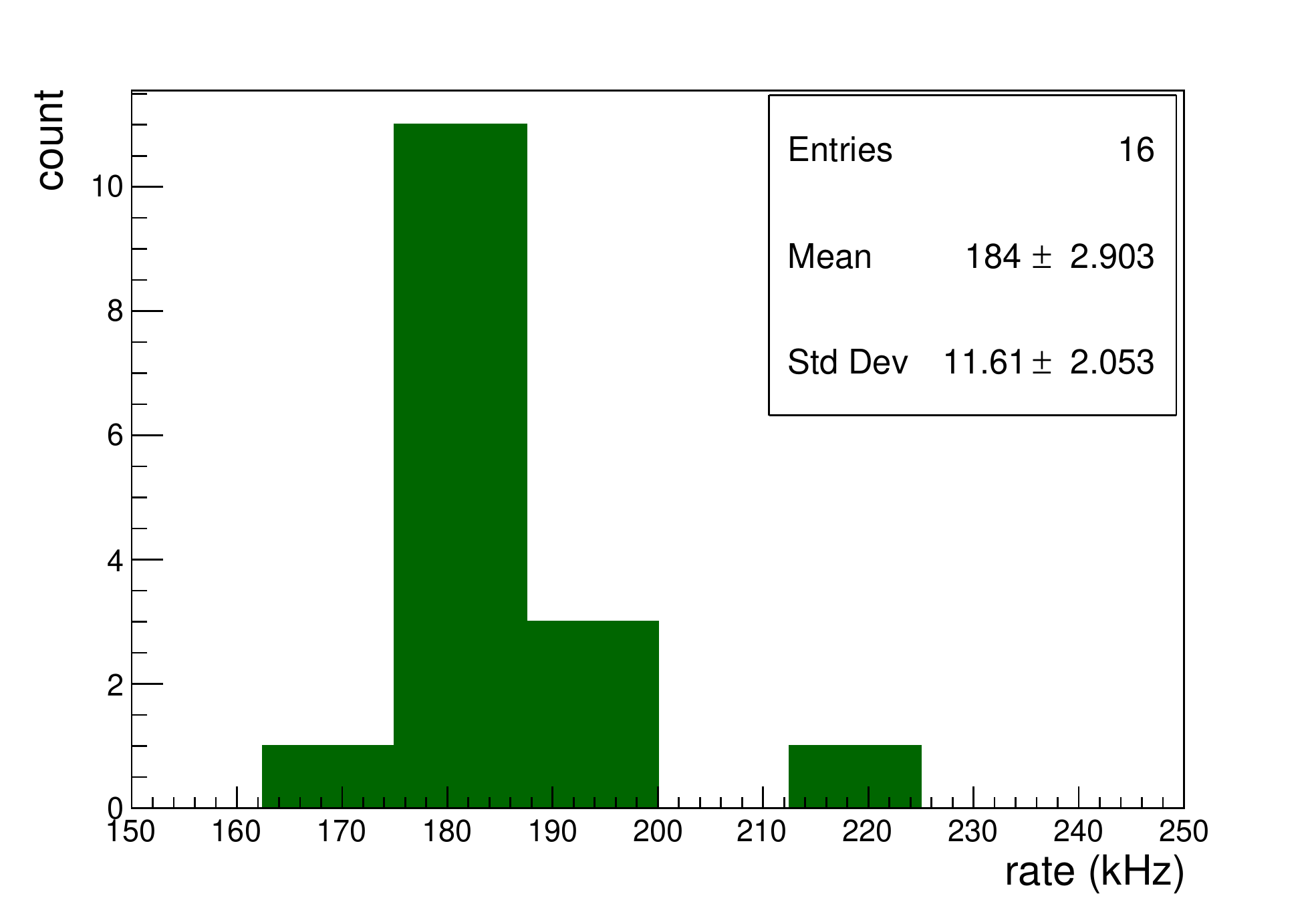}		
	\caption{Distribution of gain (top), energy resolution (middle) and count rate (bottom) over the scanned 10~cm~$\times$~10~cm area of the SM triple GEM chamber at a HV of -5075~V. The $\Delta V$ across each of the GEM foil is $\sim$~402.7 V (colour online).}\label{fig5}
\end{center}
\end{figure}
The variation in gain and count rate is found to be $\sim$~10\% and the variation in energy resolution is $\sim$~15\%. 

\begin{figure}[htb!]
\begin{center}
	\includegraphics[scale=0.40]{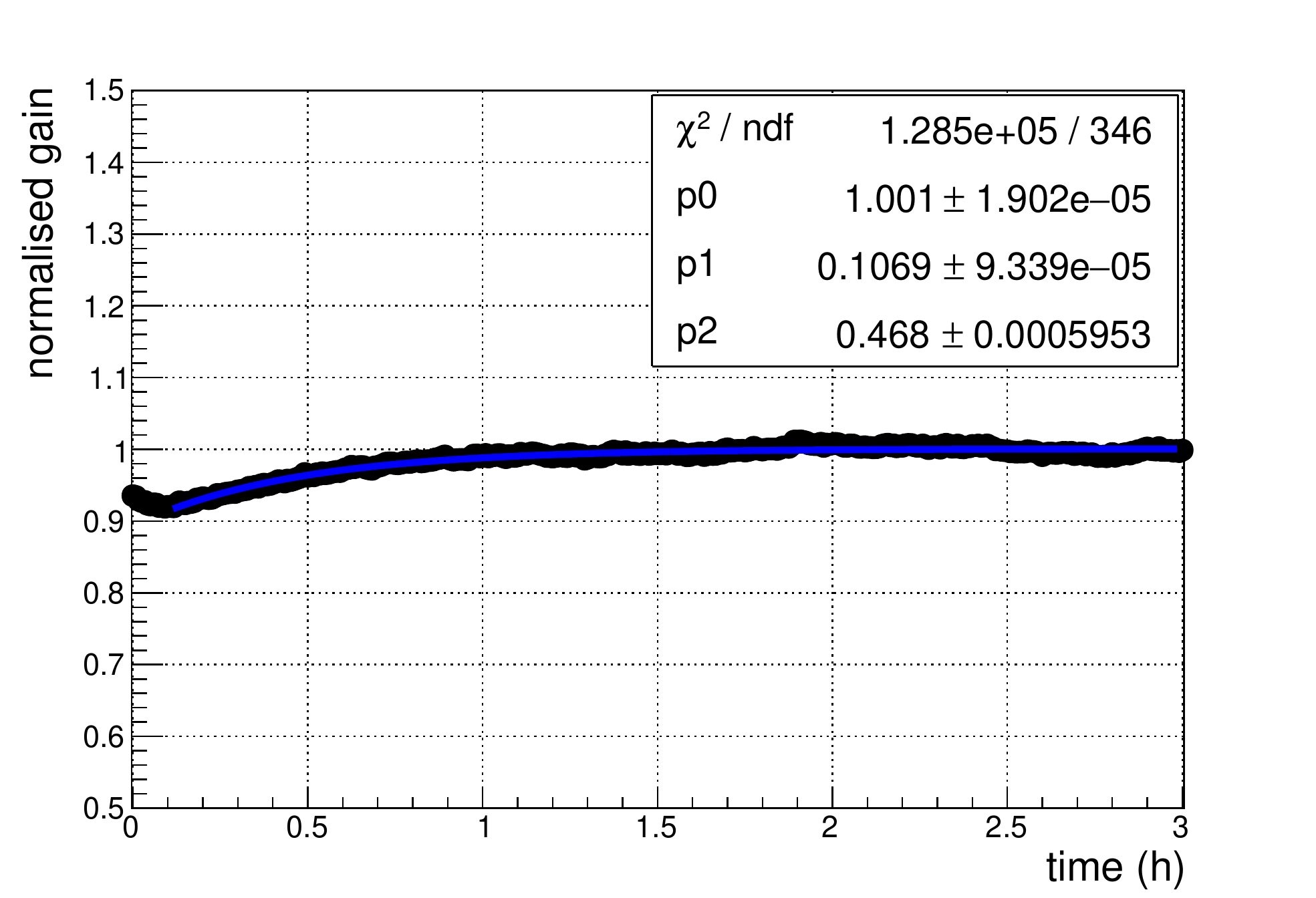}
	\includegraphics[scale=0.325]{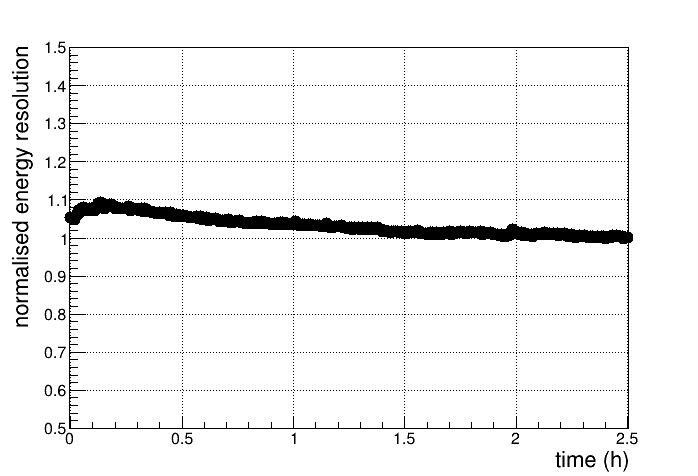}
	\caption{Variation of normalised gain~(top) and energy resolution~(bottom) as a function of time with HV of -5075~V across each of the GEM foils. The $\Delta V$ across each of the GEM foil is $\sim$~402.7 V. The normalised gain is fitted with an exponential function~($\it p0(1-p1e^{-t/p2})$) to extract the charging-up time (p2) (colour online). }\label{fig6}
\end{center}
\end{figure}

In the second case, the HV is switched ON and the source is placed on the detector as soon as the HV is reached its specific set value. The data taking is started as soon as the source is placed on the chamber. The spectra are recorded for every 30~seconds without any interval. The data is collected and then normalised to eliminate the effect of temperature~(T=t+273, t~in~$^{\circ}$C) and pressure~(p in mbar) variations on the gain and energy resolution of the chamber. The details of the T/p normalization are discussed in Ref.~\cite{s_chatterjee_charging_up_1, s_roy_gain_calculation}. The variation of the normalised gain and energy resolution as a function of time is shown in Fig.~\ref{fig6}. The energy resolution improves with an increased gain of the chamber due to the charging-up effect.
The normalised gain is fitted using an exponential parameterization to extract the charging-up time as discussed in Ref.~\cite{s_chatterjee_charging_up_1,s_chatterjee_charging_up_2}. The fitted normalised gain is shown in Fig.~\ref{fig6}~(top).
The normalised gain decreases initially due to the polarisation of the dielectric medium~\cite{s_chatterjee_charging_up_2} and after that, the charging-up phenomena take over and the gain increases and asymptotically reaches a constant value. The behaviour of the normalised gain is explained in more detail in Ref.~\cite{s_chatterjee_charging_up_2}. To find out the variation in gain, energy resolution and count rate after the charging-up phenomena is completed, the gain and energy resolution are measured after $\sim$~150 minutes of exposure of the chamber with X-ray from Fe$^{55}$ source and the count rate is recorded simultaneously from the NIM scaler. 
\begin{figure}[htb!]
\begin{center}
	\includegraphics[height=5.50cm, width=6.0cm]{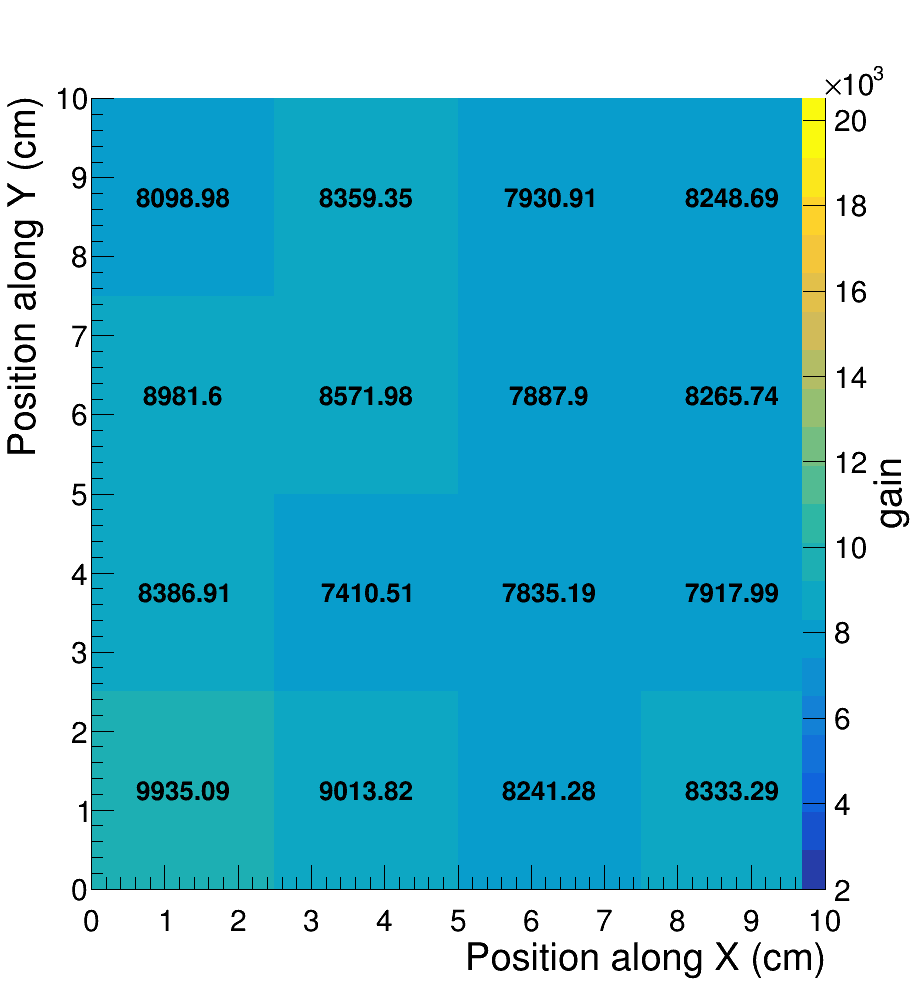}
	\includegraphics[height=5.50cm, width=6.0cm]{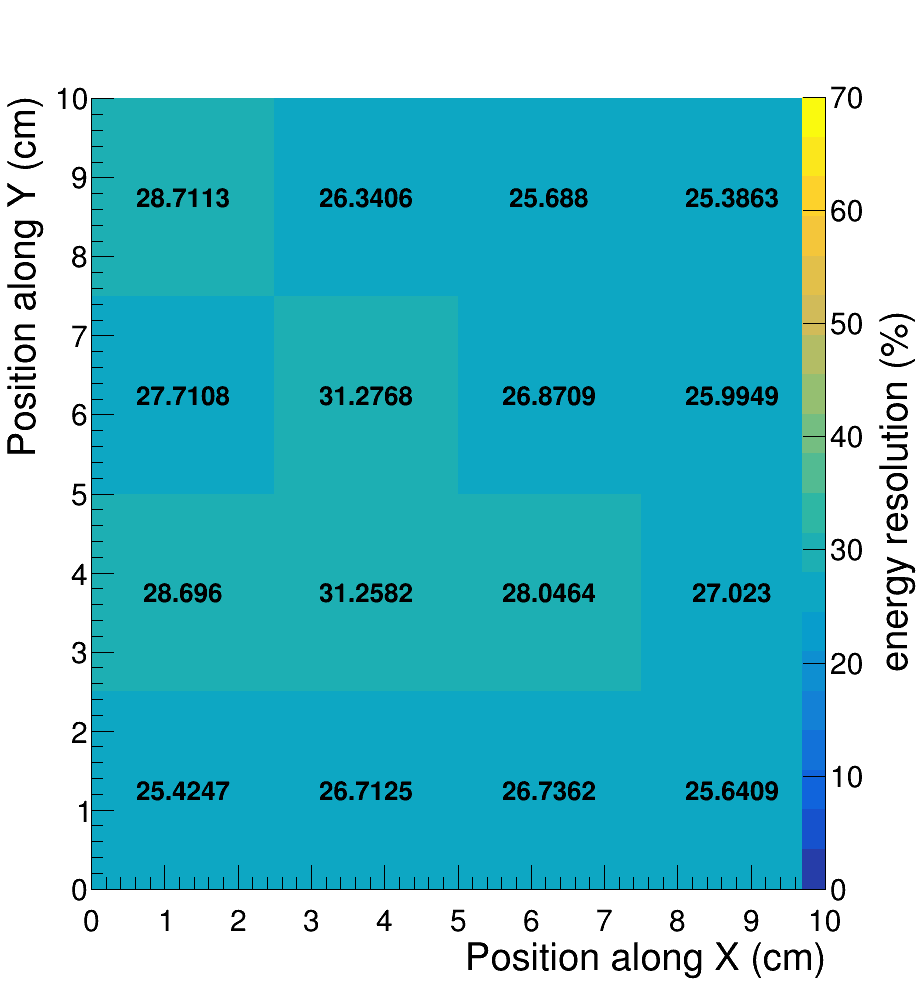}
	\includegraphics[height=5.50cm, width=6.0cm]{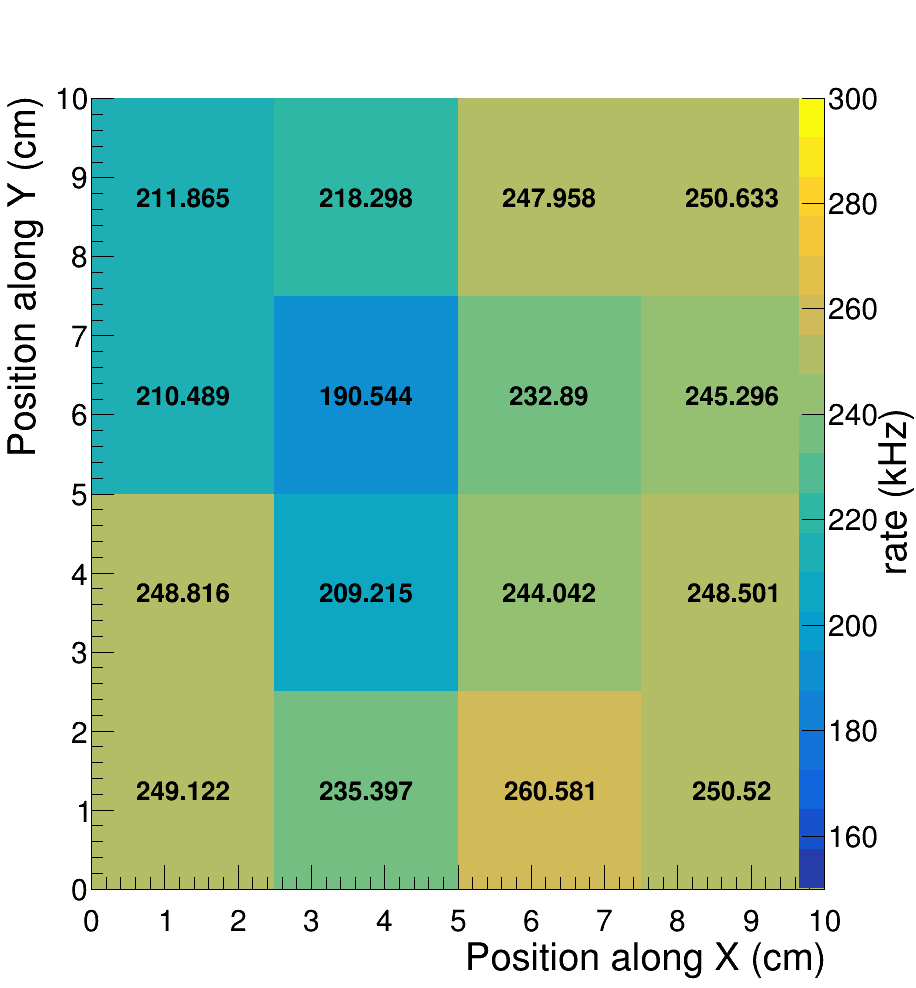}		
	\caption{Variation of gain (top), energy resolution (middle) and count rate (bottom) over the scanned 10~cm~$\times$~10~cm area of the SM triple GEM chamber with HV of -5075~V. The $\Delta V$ across each of the GEM foil is $\sim$~402.7 V (colour online). }\label{fig8}
\end{center}
\end{figure}
The variation of the gain, energy resolution and count rate over the surface with the charged-up GEM foils are shown in Fig.~\ref{fig8} and their distributions are shown in Fig.~\ref{fig9} respectively.
\begin{figure}[htb!]
\begin{center}
	\includegraphics[height=5.10cm, width=6.0cm]{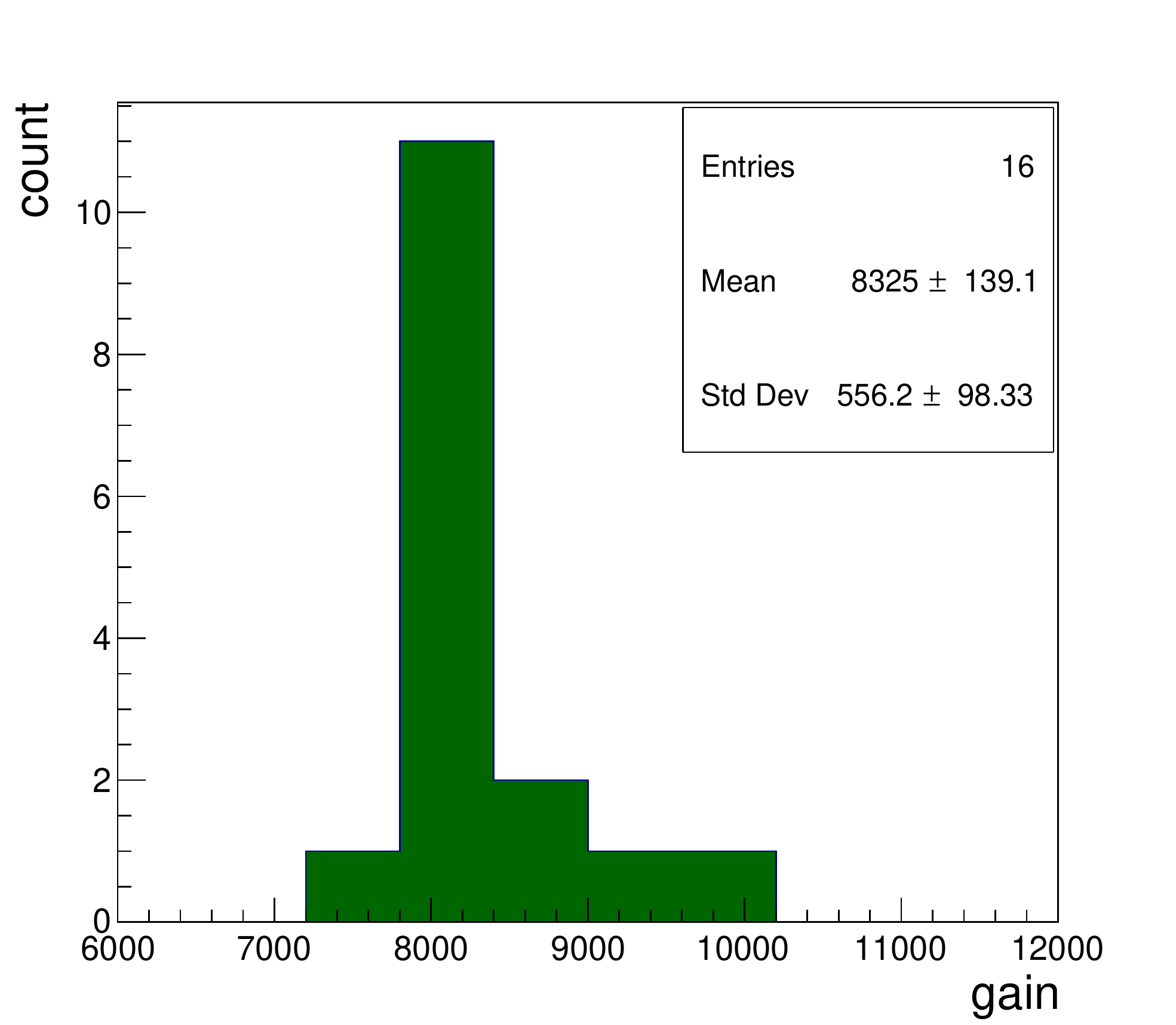}
	\includegraphics[height=5.10cm, width=6.0cm]{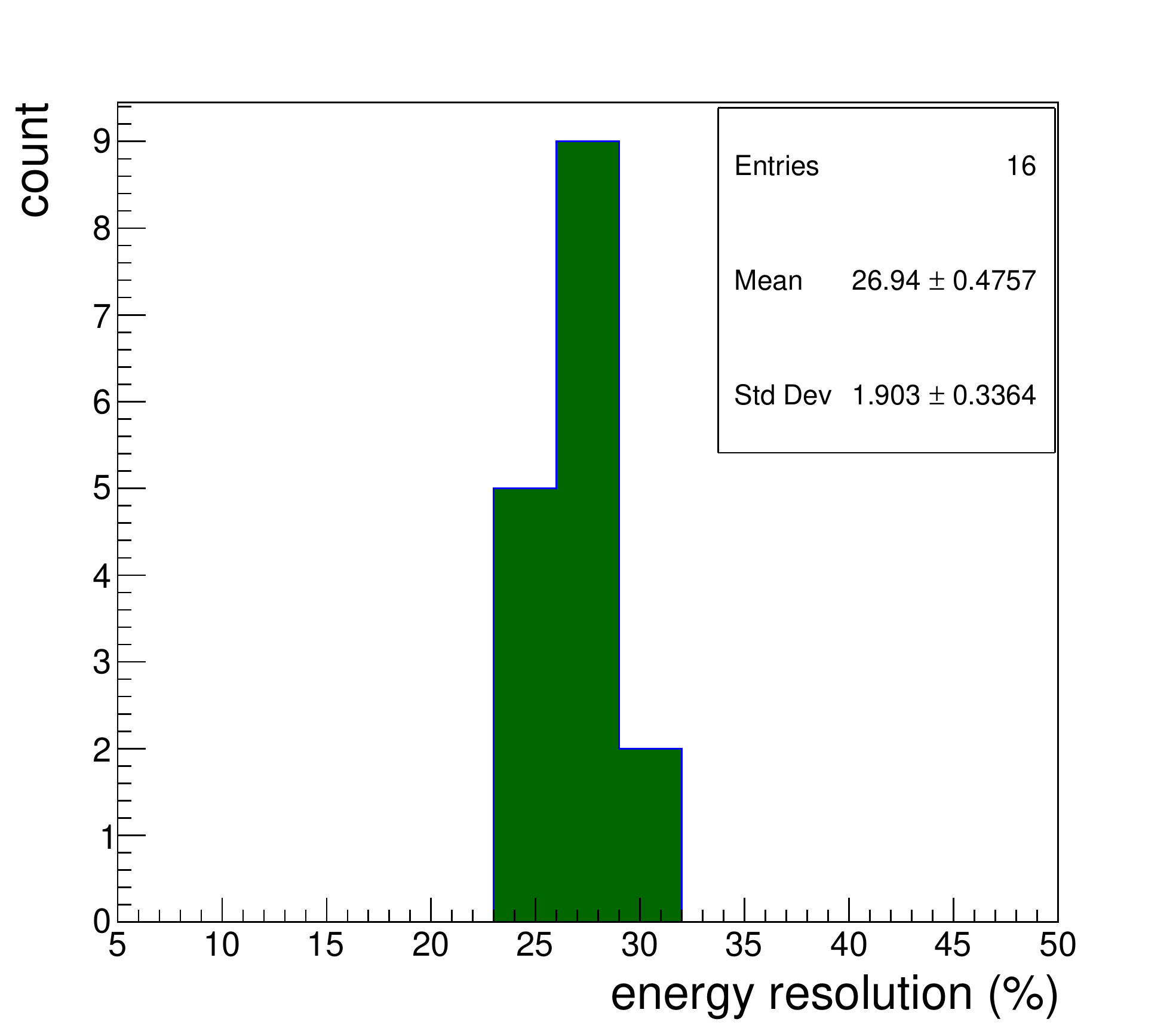}
	\includegraphics[height=5.10cm, width=6.0cm]{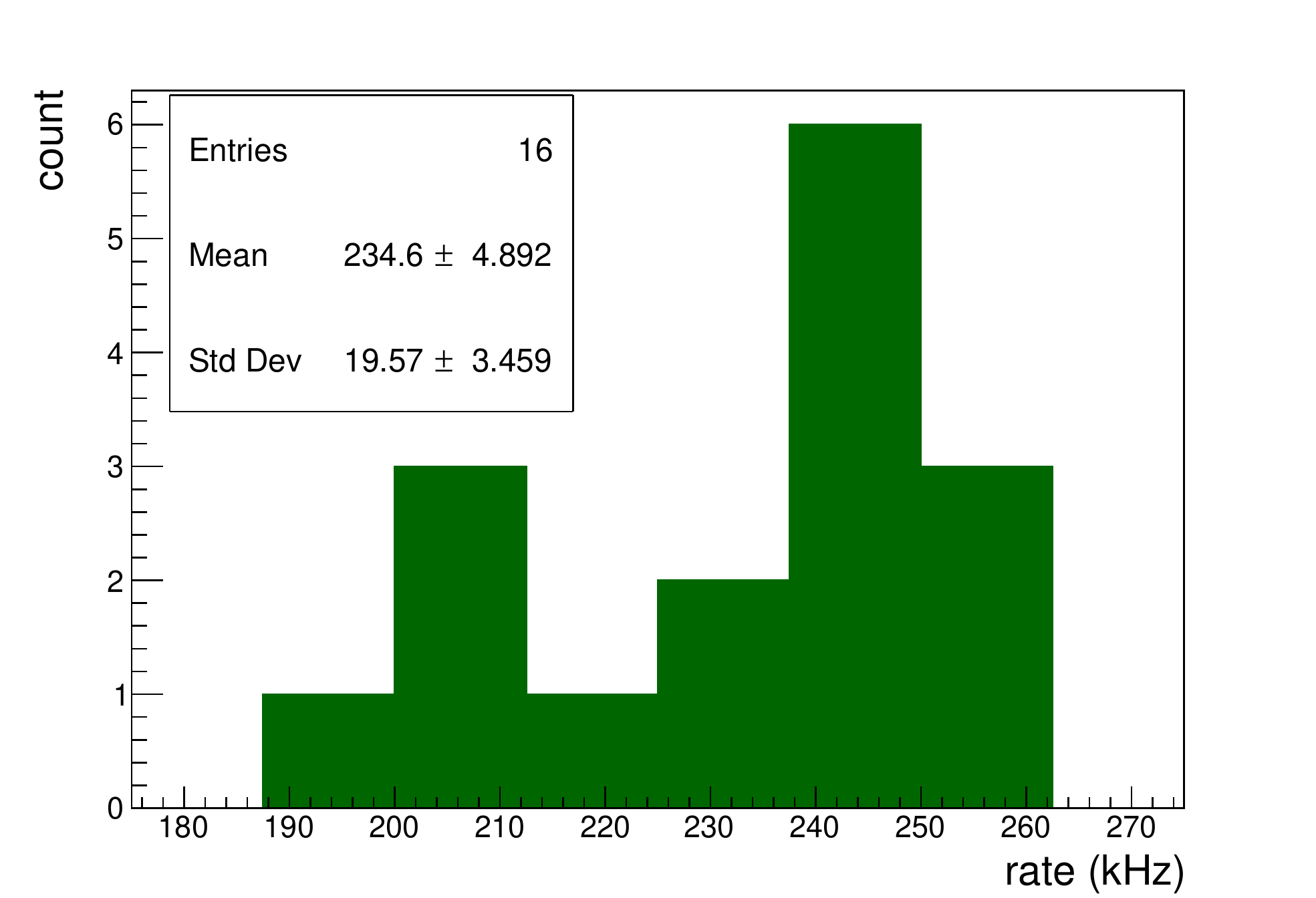}		
	\caption{Distribution of gain (top), energy resolution (middle) and count rate (bottom) over the scanned 10~cm~$\times$~10~cm area of the SM triple GEM chamber at a HV of -5075~V. The $\Delta V$ across each of the GEM foil is $\sim$~402.7 V (colour online). }\label{fig9}
\end{center}
\end{figure}
Over the scanned area, the gain, energy resolution and count rate are found to be varied $\sim$~10\%. No significant change is observed in the uniformity of the chamber over the scanned area in terms of gain, energy resolution and count rate. Only the absolute value of the gain is found to be more in the second case because of the charging-up effect of the GEM foils. Due to the increased gain, the energy resolution improves in the second case. The mean values of the gain and energy resolution without charged up GEM foils are found to be 7375~($\underline{+}171.5$), 33.31~($\underline{+}1.08$)~\% and that with the charged-up GEM foil are found to be 8325~($\underline{+}139.1$) and 26.94~($\underline{+}0.48$)~\% respectively. The count rate from the chamber is also found to be increased after the charging-up effect. The average count rate is found to be 184.0~($\underline{+}2.9$)~kHz without considering the charging-up effect and 234.6~($\underline{+}4.9$)~kHz after the charging-up effect. The ratio of the gain, energy resolution and count rate considering the charging-up phenomena is shown in Fig.~\ref{fig13} with respect to the uncharged GEM foils. The position along the x-axis indicates the sixteen different regions of the chamber. It is observed that the gain and count rate of the chamber with the charged-up GEM foils are always greater than that of the uncharged GEM foils and also the energy resolution of the chamber improves with the charging-up effect.   
\begin{figure}[htb!]
\begin{center}
	\includegraphics[height=6.0cm, width=8.0cm]{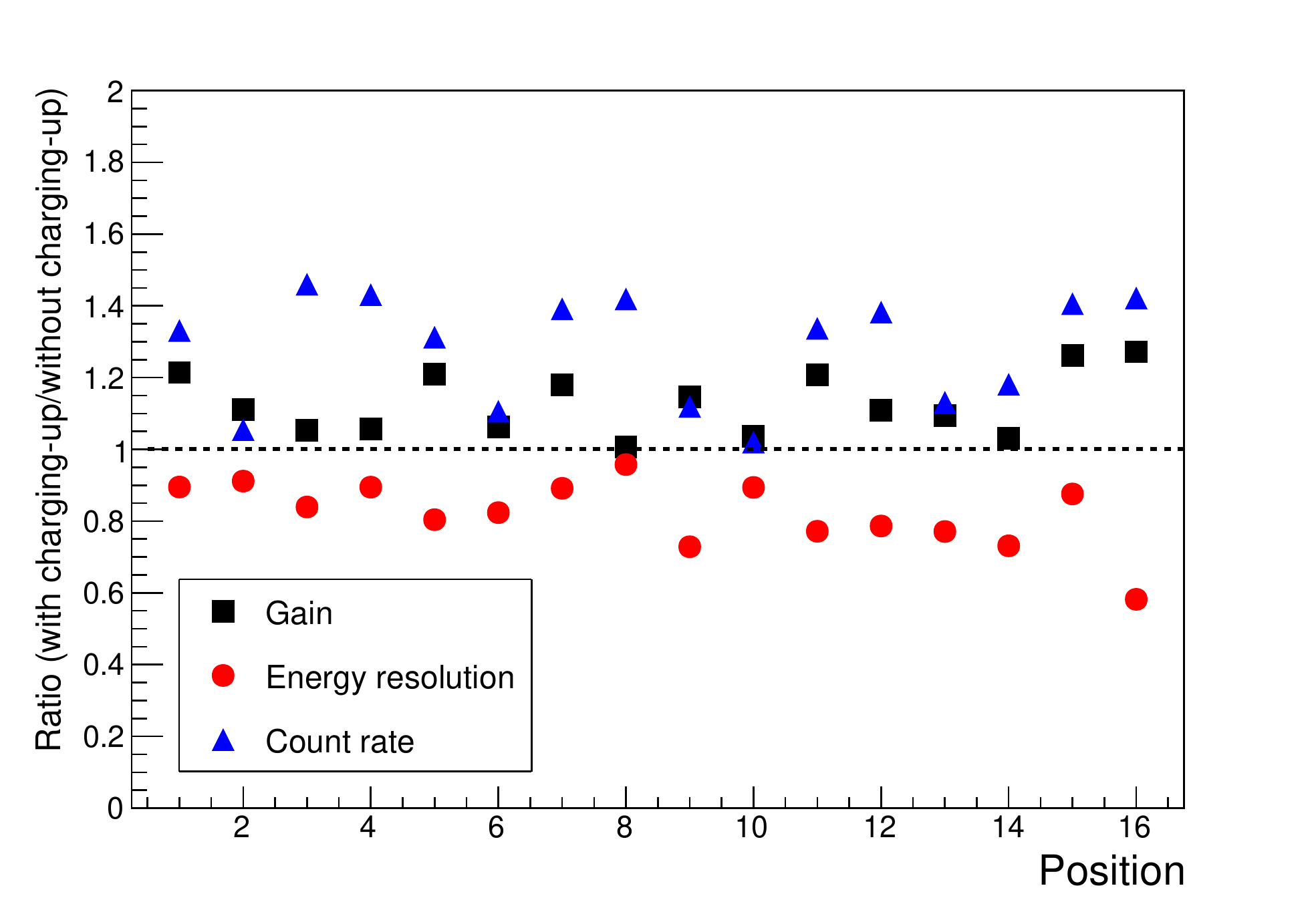}
	\caption{Ratio of gain, energy resolution and count rate with and without considering the charging-up effect of the SM triple GEM chamber at the sixteen different positions of the chamber (colour online). }\label{fig13}
\end{center}
\end{figure}


\section{Summary and Outlook}\label{summary}

The influence of the charging-up effect on the uniformity of the performance of a SM triple GEM chamber of dimension 10~cm~$\times$~10~cm is investigated using a Fe$^{55}$ X-ray source with Ar/CO$_2$ gas mixture in the 70/30 volume ratio. It is observed that with the uncharged GEM foils the variation in gain and count rate is found to be $\sim$~10\% and that in energy resolution is found to be $\sim$~15\%. The observed variation in gain, energy resolution and count rate is found to be $\sim$~10\% with the charged-up GEM foils. The mean value of the gain with and without charging-up of GEM foils are found to be 8325~($\underline{+}139.1$) and 7375~($\underline{+}171.5$) respectively. The energy resolution of the chamber also improves with increasing gain and as a result, the mean value of the energy resolution is smaller in charged-up GEM as compared with the uncharged GEM. The mean values of the energy resolution with and without charged-up GEM foils are found to be  26.94~($\underline{+}0.48$)~\% and 33.31~($\underline{+}1.08$)~\% respectively. The count rate of the chamber also increases after the charging-up of the GEM foils. The mean value of the count rates from the chamber with and without the charging-up of the GEM foils are found to be 184.0~($\underline{+}2.9$)~kHz and 234.6~($\underline{+}4.9$)~kHz respectively. Since a Fe$^{55}$ radioactive source is used to irradiate the chamber and as it emits X-rays at a fixed rate therefore we can say that due to the charging-up of the GEM foils the efficiency of the chamber also improves because of the increased gain. The probable reason behind observing less variation with the charged-up GEM foil in energy resolution could be the stabilisation of the chamber. As expected, the uniformity of the chamber doesn't depend on the charging-up of the di-electric but rather depends on the geometry of the foils and their relative separations. 

\section{Acknowledgements}

The authors would like to thank the RD51 collaboration for the support in building and initial	testing of the chamber in the RD51 laboratory at CERN. We would also like to thank Mr. Subrata Das for helping in building of the collimators used in this study. 
This work is partially supported by the research grant SR/MF/PS-01/2014-BI from DST, Govt. of India, and the research grant of the CBM-MuCh project from BI-IFCC, DST, Govt. of India. S. Biswas acknowledges the support of the DST-SERB Ramanujan Fellowship (D.O.No. SR/S2/RJN-02/2012).


\noindent

\begin{thebibliography}{50}
\bibitem{sauli_GEM}
F. Sauli, Nucl. Instrum. Methods Phys. Res. A 386 (1997) 531

\bibitem{gem_review}
A.F. Buzulutskov, Instrum Exp Tech 50, (2007) 287

\bibitem{ketzer}
B. Ketzer \textit{et al.,} Nucl. Instrum. Methods Phys. Res. A 535 (2004) 314

\bibitem{CMS_upgrade}
D. Abbaneo \textit{et al.,} Nucl. Instrum. Methods Phys. Res. A 718 (2013) 383

\bibitem{alice_upgrade}
B. Ketzer for the GEM-TPC and ALICE TPC Collaborations, Nucl. Instrum. Meth. A 732 (2013) 237


\bibitem{cbm_detector_system}
T. Balog, J. Phys. Conf. Ser. 503 (2014) 012019

\bibitem{GEM_foil}
Oliveira \textit{et al.,} United States Patent, Patent No.: US 8,597,490 B2

\bibitem{charging_up_philip}
P. Hauer \textit{et al.,} Nucl. Instrum. Methods Phys. Res. A 976 (2020) 164205

\bibitem{charging_up_azmoun}
B. Azmoun \textit{et al.,} IEEE Nuclear Science Symposium Conference Record VOL. 6 (2006) 3847

\bibitem{charging_up_alfonsi}
M. Alfonsi, Nucl. Instrum. Methods Phys. Res. A	671 (2012) 6

\bibitem{s_chatterjee_charging_up_1}
S. Chatterjee \textit{et al.,} J. Instrum. 15 (2020) T09011

\bibitem{s_chatterjee_charging_up_2}
S. Chatterjee \textit{et al.,} Nucl. Instrum. Methods Phys. Res. A 1014 (2021) 165749

\bibitem{uniformity_1}
S. Chatterjee \textit{et al.,} Nucl. Instrum. Methods Phys. Res. A 936 (2019) 491

\bibitem{uniformity_2}
M. Gola \textit{et al.,} Nucl. Instrum. Methods Phys. Res. A 951  (2020) 162967

\bibitem{uniformity_3}
R. N. Patra \textit{et al.,} Nucl. Instrum. Methods Phys. Res. A 862 (2017) 25

\bibitem{preamplifier}
CDT CASCADE Detector Technologies GmbH, Germany,\\ 
www.n-cdt.com

\bibitem{s_roy_gain_calculation}
S. Roy \textit{et al.,} Nucl. Instrum. Methods Phys. Res. A	936 (2019) 485

\end{thebibliography}
\end{document}